\documentclass[%
 reprint,
 amsmath,amssymb,
 aps,
 prd, 
 floatfix,
 nofootinbib,
 superscriptaddress]{revtex4-2}
\usepackage[
  debug=false,          
  lineno=false          
]{fixes}                
\usepackage{graphicx}   
\usepackage{xcolor}     
\usepackage{dutchcal}   
\usepackage{comment}
\usepackage{multirow}
\usepackage{hhline}
\usepackage{fontawesome}
\usepackage[normalem]{ulem}

\usepackage[
  colorlinks=true,  
  allcolors={[HTML]{2E3092}}, 
  pdfborder={0 0 0} 
]{hyperref}         
\usepackage[capitalise]{cleveref}
\usepackage{orcidlink}

\newcommand{\SG}[1]{{\color[RGB]{170,75,34}#1}}

\begin{document}

\title{Polarisation Singularities of Gravitational Waves}

\author{Claire \surname{Rigouzzo}\,\orcidlink{0000-0002-1551-4018}}
\email{claire.rigouzzo@kcl.ac.uk}
\affiliation{Department of Physics, King's College London, Strand, London WC2R 2LS, UK}

\author{Sebastian \surname{Golat}\,\orcidlink{0000-0003-3947-7634}}
\email{sebastian.1.golat@kcl.ac.uk}
\affiliation{Department of Physics, King's College London, Strand, London WC2R 2LS, UK}
\affiliation{London Centre for Nanotechnology}

\author{Alex J. \surname{Vernon}\,\orcidlink{0000-0002-3741-4202}}
\affiliation{Donostia International Physics Center (DIPC), Donostia-San Sebasti\'an 20018, Spain}

\author{Kyan \surname{Louisia}}
\email{kyan.louisia@kcl.ac.uk}
\affiliation{Department of Physics, King's College London, Strand, London WC2R 2LS, UK}
\affiliation{London Centre for Nanotechnology}

\author{Eugene \surname{Lim}\,\orcidlink{0000-0002-6227-9540}}
\affiliation{Department of Physics, King's College London, Strand, London WC2R 2LS, UK}

\author{Francisco J. \surname{Rodr\'iguez-Fortu\~no}\,\orcidlink{0000-0002-4555-1186}}
\affiliation{Department of Physics, King's College London, Strand, London WC2R 2LS, UK}
\affiliation{London Centre for Nanotechnology}

\begin{abstract}
    {Departure from idealised plane waves gives rise to intricate geometric structures in wave fields. One such structure is the polarisation singularity, which emerges when multiple monochromatic waves interfere (such as would be the case for stochastic backgrounds), producing loci of purely circular or linear polarisation. In this work, we extend the theory of polarisation singularities to gravitational waves and higher spin fields. Building on the electromagnetic description, we formulate the gravitational analogue of polarisation singularities and show that they are generic features of gravitational waves. Their dimension, however, depends on the spin of the field. We illustrate these results with simulations of plane-wave interference and analyse the resulting singularity densities \href{https://github.com/KZL358/Polarisation-Singularities}{\textcolor{black}{\faGithub}}.}
\end{abstract}
 
\maketitle

\section{Introduction}\label{sec:intro}
The direct detection of gravitational waves (GWs) has opened a new observational window on the Universe \cite{LIGO}. Beyond enabling the discovery and characterisation of compact-binary mergers and other sources (see e.g. \cite{Mandel:2018hfr,LIGOScientific:2017ycc,LIGOScientific:2020kqk,KAGRA:2021duu,KAGRA:2021vkt,Galaudage:2021rkt}), the increasing event rate \cite{GRACE} motivates statistical descriptions of GW fields that go beyond idealised settings, for example in superpositions of multiple signals or in stochastic backgrounds. 

In this work we address the deceptively straightforward setting in which multiple GWs interfere and form spatial wave structures. Structured waves, which are formed of more than one plane wave, are in fact a central topic in other disciplines \cite{Bliokh2023}, and particularly optics \cite{Rubinsztein-Dunlop2017} where fundamental theoretical and experimental studies have progressed for many decades.
Structured light beams can carry spin and orbital angular momentum \cite{Torres,Andrews}, drive complex local light-matter interactions \cite{Forbes2021,Mayer2024}, and have motivated the design of advanced metamaterials \cite{Dorrah2022}.
Perhaps at their most complex, light waves can support extraordinary features like as knots \cite{PhysRevLett.111.150404,Knots,e6e9247d3ab54f1a80459d871decfe3a} and polarisation M\"obius and Lissajous figures \cite{FREUND20101,Bauer2015,PhysRevLett.117.013601,Galvez2017,Pisanty_2019}. Applications of structured light are broad---superresolution microscopy \cite{Hell1994, Balzarotti2017}, communications and informatics \cite{Wang2012TerabitFD, Bozinovic2013, Nelson2007}, and biophysics \cite{Bustamante2021}, are but a few.

Look closely at a structured light wave---be it a vortex beam, or an arbitrary plane wave superposition---and one finds a ``skeleton" of simple, thread-like topological features that sculpt the global phase and polarisation structure of the field.
For scalar waves these features are phase singularities \cite{Nye1974} while for vector waves, these are polarisation singularities \cite{Nye1987,Berry2001,PhysRevLett.102.033902}, the focus of this work.
Light's polarisation state can be represented by a polarisation ellipse traced by the instantaneous electric-field vector over one cycle.
In structured light, this ellipse varies from point to point and has ill-defined properties at special locations: collapsing to a perfect circle (circular polarisation) or to a line segment (linear polarisation) along one dimensional strands called C lines and L lines respectively.
Most astonishingly of all, phase and polarisation singularities are demonstrably stable \cite{2015PhRvA.92f3819L}, \textit{universal} features of any linear wave field, having also been identified in acoustic and water waves \cite{Bliokh2021,Muelas_Hurtado_2022}.

Since electromagnetic waves and GWs are both massless radiative fields with two propagating degrees of freedom (for a review, see e.g \cite{Barnett2014}), it is natural to ask how these concepts extend to gravity, and what changes once the field carries spin 2 rather than spin 1. Here ``spin’’ refers to the rotational symmetry of the field: a spin-1 configuration returns to itself only after a full rotation, while a spin-2 configuration repeats after half a rotation. 

In this letter we extend the polarisation singularity framework to gravitational waves\footnote{Note that a different concept of topological singularities for higher spins was explored in \cite{dennis2001topological}, but their physical interpretation was not discussed.}. We show that circular-polarisation singularities remain generically one-dimensional curves in three-dimensional space, whereas linear-polarisation singularities become isolated point defects for gravitational waves. This result holds for any higher spin wave. Finally, we illustrate these predictions with simulations of random plane-wave interference, and estimate the corresponding singularity densities. We also comment on the GW phenomenology of such singularities.
\section{Polarisation singularities for electromagnetic waves} \label{sec:pola_ENM}
\begin{figure}[t!]
    \centering
    \includegraphics[width=\columnwidth]{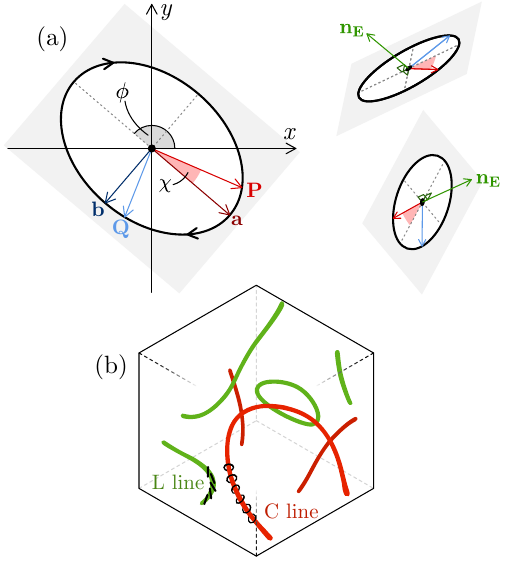}
    \caption{Generic electric field polarisation.
    (a) left: a polarisation ellipse in the $xy$ plane described by $\mathbf{E}=\mathbf{P}+i\mathbf{Q}$, and its semi-major and semi-minor axes aligned with the real vectors $\mathbf{a}$ and $\mathbf{b}$.
    (a) right: polarisation ellipses with arbitrary 3D orientation determined by the normal to the plane of the ellipse, $\mathbf{n}_\mathbf{E}$.
    (b): generic depiction of red C lines and green L lines in a volume of space containing an electromagnetic field (e.g., random 3D plane wave interference).}
    \label{ellipse}
\end{figure}

In this section, we briefly review polarisation singularities in monochromatic electromagnetic waves, following closely the seminal paper \cite{Nye1987}. At each point $\mathbf r$, the instantaneous electric field $\boldsymbol{\mathcal{E}}(\mathbf r,t)$ is periodic in time with angular frequency $\omega$, and
over one optical cycle, $\boldsymbol{\mathcal{E}}(\mathbf r,t)$ traces a planar curve. In the generic case, this curve is an ellipse, with linear and circular polarisations arising as degenerate limiting cases.
The trajectory of $\boldsymbol{\mathcal{E}}(\mathbf{r},t)$, and therefore the ellipticity and orientation of the electric polarisation ellipse (\cref{ellipse}(a)), is completely determined by the real and imaginary parts of the electric field phasor $\mathbf{E}(\mathbf{r})=\mathbf{P}(\mathbf{r})+i\mathbf{Q}(\mathbf{r})$, which is defined by:
\begin{equation}
    \boldsymbol{\mathcal{E}}(\mathbf{r},t)=\operatorname{Re}\!\left[\mathbf{E}(\mathbf{r})e^{-i\omega t}\right]\!=\mathbf{P}(\mathbf{r})\cos{\omega t}+\mathbf{Q}(\mathbf{r})\sin{\omega t}\,.
    \label{pola_ellipse}
\end{equation}
where $\mathbf{P}$ and $\mathbf{Q}$ are real vector fields, and we now omit explicit $\mathbf{r}$-dependence of quantities.\footnote{Throughout, calligraphic type, $\boldsymbol{\mathcal{E}}(\mathbf{r},t)$, is used for real-valued fields, whereas standard Roman type, $\mathbf{E}(\mathbf{r})$, denotes their complex phasor representations. The two are related by \cref{pola_ellipse}.}
Clearly, $\boldsymbol{\mathcal{E}}$ oscillates in the plane that contains $\mathbf{P}$ and $\mathbf{Q}$, and the normal to this plane is
\begin{equation}\label{normal}
    \mathbf{n}_\mathbf{E}=\tfrac12\operatorname{Im}(\mathbf{E}^*\!\times\mathbf{E})=\mathbf{P}\times\mathbf{Q}\,.
\end{equation}
{Now, multiplying the complex vector $\mathbf{E}=\mathbf{P}+i\mathbf{Q}$ by a global phase factor $e^{-i\chi}$ does not affect the shape of the ellipse, it merely shifts the time origin by $\chi/\omega$.} This phase freedom allows us to select a ``rectifying’’ phase $\chi$ such that $e^{-i\chi}\mathbf{E}=\mathbf{a}+i\mathbf{b}$ where the real vectors $\mathbf{a}$ and $\mathbf{b}$ align with the semi-major and semi-minor axes of the polarisation ellipse (see \cref{ellipse}).
The phase required for this is $\chi=\arg(\psi_\mathbf{E})/2$, where
\begin{equation}\label{psi_EM}
    \psi_\mathbf{E}=\mathbf{E}\cdot \mathbf{E}=||\mathbf{P}||^2-||\mathbf{Q}||^2+2 i\mathbf{P}\cdot \mathbf{Q}\,.
\end{equation}
A field phasor for which one of \cref{normal} or \eqref{psi_EM} is zero leads to a special configuration of the polarisation ellipse: circular or linear polarisation. Circular and linear trajectories of $\boldsymbol{\mathcal{E}}$ are known as \emph{polarisation singularities}, as in each case a different characteristic of the generic polarisation ellipse is undefined. Having zero values in an otherwise smoothly varying vector $\mathbf{n}_\mathbf{E}(\mathbf{r})$ or scalar $\psi_\mathbf{E}(\mathbf{r})$ field has topological implications in the surrounding space \cite{Vernon:23}.
We will now show that in a generic monochromatic field, one expects \textit{lines} of circular and linear polarisation, forming our polarisation singularities.

\paragraph*{C lines}
arise at points where the electric field vector traces a perfect circle. This occurs when $\mathbf{P} \perp \mathbf{Q}$ ($\operatorname{Im}\psi_\mathbf{E}=0$) and $||\mathbf{P}||=||\mathbf{Q}||$ ($\operatorname{Re}\psi_\mathbf{E}=0$), i.e., the polarisation ellipse is degenerate and has no semi-major or semi-minor axes, which means that the \emph{rectifying angle}, $\chi=\arg(\psi_\mathbf{E})/2$, becomes undefined. As shown in \cref{app_codimension}, two independent real conditions imposed on points in three-dimensional space generically lead to one-dimensional curves: $3-2=1$. Loci of purely circular polarisation are therefore aligned in C lines. A compact way to express the same conditions is through $\psi_\mathbf{E}=\mathbf{E}\cdot \mathbf{E}=0$. An example is shown as a red curve in \cref{ellipse}(b).

\paragraph*{L lines}
arise at points where the polarisation is linear, so that the field oscillates along a straight line. This occurs when $\mathbf{P}$ and $\mathbf{Q}$ are parallel (or one of them vanishes), so that the polarisation ellipse collapses into a line segment. This makes the normal to the plane of the ellipse undefined, $\mathbf{n}_\mathbf{E}=\mathbf{P}\times \mathbf{Q}=\mathbf{0}$. Although this might appear to impose three independent conditions, it is important to consider the underlying degrees of freedom. A pair of real vectors $(\mathbf{P},\mathbf{Q})$ in three dimensions contains $3+3=6$ degrees of freedom. Requiring $\mathbf{P}\parallel\mathbf{Q}$ means that there exists a real scalar $\alpha$ such that $\mathbf{Q}=\alpha\mathbf{P}$, and the pair $(\mathbf{P},\alpha)$ carries only $3+1=4$ degrees of freedom. Hence, the requirement that the two vectors share a direction provides \emph{only} two independent constraints. Consequently, the set of points where the polarisation is exactly linear forms {$3-2=1-$dimensional curves in $3$D space}---the L lines. An example is shown as a green curve in \cref{ellipse}(b).

These polarisation singularities are topologically protected \cite{Spaegele2023}: to perturb the field, i.e., to add a small but otherwise arbitrary complex vector field $\delta\mathbf{E}$ to $\mathbf{E}$, in general, fails to destroy its polarisation singularities, merely displacing them from their original position \cite{2015PhRvA.92f3819L}. If this is done by continuously varying a parameter $\epsilon$ in $\mathbf{E}\mapsto\mathbf{E}+\epsilon\delta\mathbf{E}$ between zero and one, the deformation of the lines will be smooth.

Both \emph{C~lines} and \emph{L~lines} are generic features of monochromatic waves, permeating focussed beams of light and even superpositions of randomly polarised, randomly propagating plane waves. While the electric and magnetic fields of structured light can be polarised differently ($\boldsymbol{\mathcal{E}}$ and $\boldsymbol{\mathcal{B}}$ sweep different ellipses), the electric field tends to take precedence in optics because most detectors or materials that might interact with light are relatively insensitive to magnetic polarisation \SG{\cite{Vernon2025Jul,Golat2025Oct}}. Last but not least, we note that the ellipse traced by the vector potential $\boldsymbol{\mathcal{A}}$ in the radiation gauge ($\boldsymbol\nabla\cdot\boldsymbol{\mathcal{A}}=0$) is merely a scaled version of the electric-field ellipse. For a monochromatic wave, $\mathbf{E}=i\omega \mathbf{A}$; 
hence, the C line condition written in terms of $\mathbf{E}$ is identical to that written in terms of $\mathbf{A}$, $\psi_\mathbf{E}=-\omega^2\psi_\mathbf{A}=0$, and similarly, the L line conditions coincide, $\mathbf{n}_\mathbf{E}=\omega^2\mathbf{n}_\mathbf{A}=\mathbf{0}$.

\section{Polarisation singularities for gravitational waves} \label{sec:pola_GW}

\begin{figure}[t!]
    \centering
    \includegraphics[width=\columnwidth]{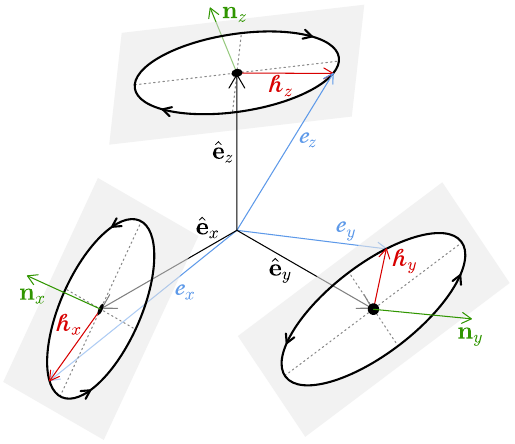}
    \caption{Generic gravitational field polarisation at a single point $\mathbf{r}$. A gravitational wave displaces the local orthonormal basis vectors $\hat{\mathbf e}_i$ by $\boldsymbol{\mathcal h}_i(\mathbf r,t)=\tfrac12\mathcal{h}^a{}_i(\mathbf r,t)\hat{\mathbf e}_a$, which traces an ellipse in time. The displacement vectors $\boldsymbol{\mathcal h}_i$ are shown in red, and the normal to each ellipse, $\mathbf n_i$, is shown in green.}
    \label{ellipseGR}
\end{figure}

Building on the electromagnetic case, we now explore polarisation singularities in gravitational waves. In close analogy with the radiation-gauge vector potential $\mathcal{A}_i(\mathbf{r}, t)$, the radiative degrees of freedom of linearised gravity are encoded in the metric perturbation $\mathcal{h}_{ij}(\mathbf{r}, t)$ in the transverse–traceless (TT) gauge ($\partial_i\mathcal{h}^{i}{}_j=\mathcal{h}^i{}_i=0$).\footnote{Here the latin indices $i,j,k,\dots$ label components in the coordinate basis (e.g.\ $x,y,z$), i.e. any spatial vector can be decomposed into $\mathbf{v} = v^i \hat{\mathbf{e}}_i$, where $\hat{\mathbf{e}}_i$ form an orthonormal basis satisfying $\hat{\mathbf{e}}_i\cdot\hat{\mathbf{e}}_j=\delta_{ij}$. A repeated index appearing once up and once down is implicitly summed, e.g.
$A_{ij}B^i = A_{xj}B_x + A_{yj}B_y + A_{zj}B_z$.}
The monochromatic wave can also be fully described using a time-independent complex, symmetric, trace-free tensor $h_{ij}(\mathbf{r})=P_{ij}(\mathbf{r})+iQ_{ij}(\mathbf{r})$:
\begin{equation}
\begin{split}
    \mathcal{h}_{ij}(\mathbf{r}, t)&=\operatorname{Re}[h_{ij}(\mathbf{r})e^{-i\omega t}]\\&=P_{ij}(\mathbf{r})\cos{\omega t}+Q_{ij}(\mathbf{r})\sin{\omega t}\;.
\end{split}
\end{equation}
This equation has a very similar structure to \cref{pola_ellipse}, except
that the fields $P_{ij}$ and $Q_{ij}$ are rank-2 spatial tensors  rather than rank-1 vectors, which makes finding the local polarisation ellipse at a point $\mathbf{r}$ less straightforward. However, we can make the analogy with electromagnetism manifest by noting that the metric $g_{ij}$ can be constructed out of the inner product of two local basis vectors $\boldsymbol{e}_i(\mathbf{r}, t)$
as follows \cite{MTW}:
\begin{equation}\label{metric}
    \boldsymbol{e}_i(\mathbf{r}, t)\cdot\boldsymbol{e}_j(\mathbf{r}, t)=g_{ij}(\mathbf{r}, t)=\delta_{ij}+\mathcal{h}_{ij}(\mathbf{r}, t)\;,
\end{equation}
to linear order in the metric perturbation $\mathcal{h}_{ij}$.\footnote{A word on notation for the more differential geometrically inclined: unhat scripted bold variables indicate abstract vectors in the \emph{tetrad} basis $\hat{\mathbf{e}}_a$, with indices $a,b,c,\dots$ labelling this basis so that $\boldsymbol{e}_i(\mathbf{r}, t) \equiv e^a_i(\mathbf{r}, t)\hat{\mathbf{e}}_a$. The inner product operator $\cdot$ denotes contraction with $\delta_{ab}$, such that $\delta_{ab}e^{a}_ie^{b}_i = g_{ij}$. To linear order in $\mathcal{h_{ij}}$, 
    $\boldsymbol{e}_i(\mathbf{r}, t)=\hat{\mathbf{e}}_i+ \boldsymbol{\mathcal{h}}_i(\mathbf{r}, t)=[\delta^a{}_i+\tfrac12\mathcal{h}^{a}{}_{i}(\mathbf{r}, t)]\hat{\mathbf{e}}_a\, $.
}
These local basis vectors are perturbed by the gravitational wave to the linear order as follows:
\begin{equation}\label{perturbedvector}
    \boldsymbol{e}_i(\mathbf{r}, t)=\hat{\mathbf{e}}_i+ \boldsymbol{\mathcal{h}}_i(\mathbf{r}, t)=\hat{\mathbf{e}}_i+\tfrac12\mathcal{h}^{a}{}_{i}(\mathbf{r}, t)\hat{\mathbf{e}}_a\,
\end{equation}
where the vector $\boldsymbol{\mathcal{h}}_i(\mathbf{r}, t)$ now plays the analogous role to the gauge potential
$\boldsymbol{\mathcal{A}}(\mathbf{r}, t)$. Thus, each column of the metric perturbation $h_{ij}$ describes the oscillatory motion of the corresponding basis vector. Each resulting perturbation vector $\boldsymbol{\mathcal{h}}_j$ traces an ellipse in time, which can be conveniently described by the complex phasor $\mathbf{h}_j(\mathbf{r})=\mathbf{P}_{j}(\mathbf{r})+i\mathbf{Q}_{j}(\mathbf{r})$
\begin{equation}
\begin{split}
    \boldsymbol{\mathcal{h}}_j(\mathbf{r}, t)&=\operatorname{Re}\!\left[\mathbf{h}_j(\mathbf{r})e^{-i\omega t}\right]\!\\&=\tfrac12[P^{i}{}_{j}(\mathbf{r})\cos{\omega t}+Q^{i}{}_{j}(\mathbf{r})\sin{\omega t}]\hat{\mathbf{e}}_i\;,
\end{split}
\end{equation}
as can be seen in \cref{ellipseGR}.\footnote{For example, the ellipse associated with the $x$-directed basis vector has a normal $\mathbf{n}_x=\mathbf{P}_x\times\mathbf{Q}_x=\tfrac14\varepsilon^{ijk}P_{jx}Q_{kx}\hat{\mathbf{e}}_i$ and a complex scalar encoding its linear eccentricity and rectifying angle $\psi_x=\mathbf{h}_{x}\cdot\mathbf{h}_{x}=\tfrac14h_{ix}h^{i}{}_{x}$.} %
These ellipses depend on the choice of basis, as they describe the oscillatory motion of individual basis vectors. It is nevertheless possible to construct a basis-independent normal by contracting the free index, yielding:
\begin{equation}\label{normal_GR}
    \!\!\mathbf{n}_{h}=4\mathbf{P}_l\times\mathbf{Q}^l=\tfrac12\varepsilon^{ijk}\operatorname{Im}(h_{jl}^*h_{k}{}^l)\hat{\mathbf{e}}_i=\varepsilon^{ijk}P_{jl}Q_{k}{}^l\hat{\mathbf{e}}_i\,,\!\!
\end{equation}
which is proportional to the vector sum of the individual normals $\mathbf{n}_{h}=4(\mathbf{n}_{x}+\mathbf{n}_{y}+\mathbf{n}_{z})$. This vector is also proportional to the spin angular momentum density (see \cref{app:higher_spin}). In a similar manner, one may define a basis-independent complex scalar
\begin{equation}\label{psi_GR}
    \!\!\psi_{h}=4\mathbf{h}_{l}\cdot\mathbf{h}^{l}=h_{ij}h^{ij}=P_{ij}P^{ij}-Q_{ij}Q^{ij}+2iP_{ij}Q^{ij},\!\!
\end{equation}
which is likewise $\psi_{h}=4(\psi_{x}+\psi_{y}+\psi_{z})$. This scalar controls the oscillatory part of the instantaneous energy density (\cref{app:higher_spin}). However, since each basis vector traces its own ellipse in real space, these quantities do not, in general, correspond to a single polarisation ellipse whose time evolution represents the perturbation. In other words, no unique real-space ellipse with a well-defined phase can be associated with the combined motion; instead, $\mathbf n_h$ and $\psi_h$ characterise the geometry of the ensemble of ellipses rather than a single physical trajectory. \Cref{visualisation} shows that this ellipse corresponds to the orientation average of all the ellipses associated to all possible unit vectors. Imposing either the vanishing of \cref{normal_GR} or \eqref{psi_GR}, we again find that there are two families of generic polarisation singularities.

\paragraph*{C lines} arise at points where $\psi_{h}=h_{ij}h^{ij}=0$. This occurs when the real and imaginary parts of the metric perturbation are orthogonal $P_{ij}Q^{ij}=0$ ($\operatorname{Im}\psi_h=0$) and have equal norm $P_{ij}P^{ij}=Q_{ij}Q^{ij}$ ($\operatorname{Re}\psi_h=0$). At these points, the instantaneous local energy density is strictly time independent.
This property is not specific to gravitational waves: the vanishing of $\psi$ implies constant energy density for electric C lines and, more generally, for C singularities of arbitrary spin-$s$ fields, as discussed in \cref{app:higher_spin}.
In three-dimensional space, the two independent conditions defining $\psi_h=0$ generically give rise to one-dimensional singular sets, corresponding here to gravitational C lines.

\paragraph*{L points} arise at spatial positions where the vector defined in \cref{normal_GR} vanishes, $n^i_{h}=\varepsilon^{ijk}P_{jl}Q_{k}{}^l=0$, corresponding to points of zero local spin angular momentum density. This property is again universal: L singularities correspond to zero spin density even for electric fields and, more generally, for any spin-$s$ fields, as discussed in \cref{app:higher_spin}. In this tensorial setting, the condition $n_h^{i}=0$ imposes, in general, \textit{three} independent real constraints. This becomes clear when one considers the associated degrees of freedom. A pair of symmetric, traceless tensors $(P_{ij},Q_{ij})$ has $5+5=10$ degrees of freedom. The requirement $n_h^{i}=0$ is equivalent to demanding that $P_{ij}$ and $Q_{ij}$ commute,
$P_{il}Q^{lj}=P_{jl}Q^{li}$, which ensures that they can be simultaneously diagonalised. In that case, there exists a rotation $R\in\mathrm{SO}(3)$ such that $R^{\intercal} P R = \operatorname{diag}(p_1,\, p_2,\, p_3)$, where the trace $p_1+p_2+p_3$ {vanishes}, with an analogous representation for $R^{\intercal} Q R$. A simultaneously diagonalisable pair is therefore specified by the rotation $R$ (three degrees of freedom) together with the two independent diagonal components of each tensor, giving the total of $3+2+2=7$ degrees of freedom. Thus, enforcing commutation removes $10-7=3$ degrees of freedom, confirming that the condition $n_h^{i}=0$ provides three independent constraints, unlike in the electromagnetic case where it was just two. The key difference with respect to \cref{normal} is that \cref{normal_GR} is a \emph{sum} of \emph{three} cross-products. Consequently, the loci at which the normal vector vanishes are generically \emph{point-like} in three-dimensional space: one expects isolated L points rather than extended filamentary structures such as L lines in the electromagnetic case. Interestingly, each L point carries a topological index{:} a zero of a smooth real three dimensional vector field has degree $+1$ (sink or source) or $-1$ (saddle), given by how many times the unit vector wraps around the unit sphere near the zero \cite{Vernon:23}. This conservation property constrains how such zeros can appear, disappear, and evolve, as will be shown later.

\section{Simulations of polarisation singularities}
 \label{sec:Simulation}
\begin{figure}[t!]
		\centering
            \includegraphics[width=.9\columnwidth]{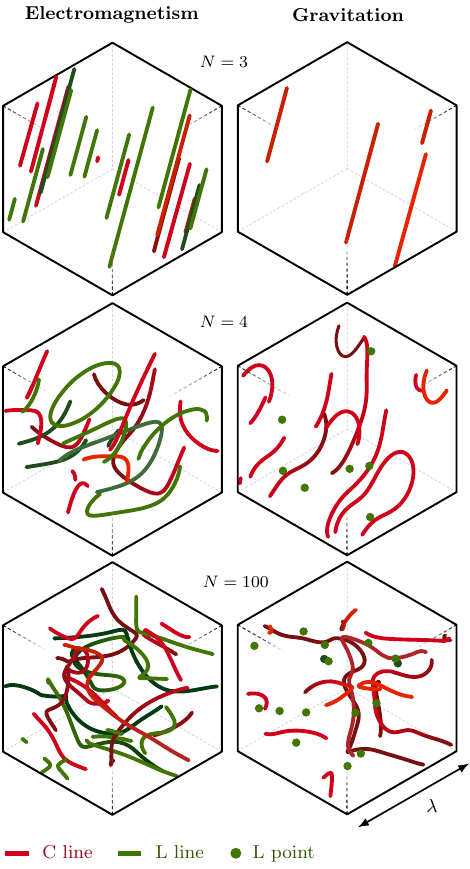}
			\caption{
            Polarisation singularities for the interference of $N=3,4,100$ random plane waves.
            Left: electromagnetic C (red) and L (green) lines.
            Right: gravitational C lines (red) and L points (green).
            {The gradient of green and red is to help show depth.} The code is available online \href{https://github.com/KZL358/Polarisation-Singularities}{\textcolor{black}{\faGithub}} \cite{figshare}. } 
			\label{interference}
	\end{figure}

Now that we have built the theory of polarisation singularities for gravitational waves, we can verify their occurrences and properties numerically. Let us consider a superposition of $N$ plane waves, all with wavelength $\lambda$,
\begin{equation}
    \!\!A_{I}({\mathbf{r}})=\sum_{n=1}^N\sum_{P}a_n^P \hat{e}^P_{I}({\mathbf{k}}_n)e^{i\mathbf{k}_n\cdot\mathbf{r}},\!\!
\end{equation}
where $A_I({\mathbf{r}})$ denotes either the electric field phasor $E_i({\mathbf{r}})$ or the metric tensor perturbation phasor $h_{ij}({\mathbf{r}})$, with $I$ representing one ($i$) or two ($ij$) indices.
The wavevectors $\smash{{\mathbf{k}}}_n$ are drawn independently and uniformly over the unit sphere, ensuring statistical isotropy, while the wavenumber magnitude is fixed to $|{\mathbf{k}}_n|=2\pi/\lambda$. The polarisation vectors $\hat{e}^P_{i}(\mathbf{k}_n)$ and symmetric traceless tensors $\hat{e}^P_{ij}(\mathbf{k}_n)$ form orthonormal bases transverse to ${\mathbf{k}}_n$, with $P$ labelling the polarisation state (see \cref{app_extension} for an explicit construction). The complex amplitudes $a_{n}^P$ are taken to be independent zero-mean circular Gaussian random variables with covariance
$\langle a_{n}^{P}a_{m}^{P'*}\rangle=(\sigma_A^2/N)\,\delta_{nm}\delta_{PP'}$. The normalisation is chosen such that the total field amplitude remains finite as $N$ increases, with $\sigma_A$ setting its overall scale.
Such systems are physically relevant as it is realised for example by considering $N$ electromagnetic dipoles next to each other, or $N$ black hole binary systems. %
The results are shown in \cref{interference} for a $\lambda\times\lambda\times\lambda$ volume, for electromagnetic and gravitational waves. We can therefore confirm the theoretical results proven in \cref{sec:pola_GW}: circular polarisation singularities are \textit{lines} for both electromagnetic and gravitational waves, whereas the linear polarisation singularities are lines for electromagnetic waves, but \textit{points} for gravitational waves. The main takeaway from \cref{interference} is that circular polarisation consists of lines, regardless of how many plane waves we sum or the nature of the wave (electromagnetic or gravitational). It is also clear from \cref{interference} that electromagnetic and gravitational wave C line geometry share the same behaviour: when $N=3$, they are aligned and straight. As soon as $N > 3$, the lines are curved and not aligned.
Note, on the other hand, that no linear polarisation is produced in the interference of three gravitational waves. We can also verify their stability by continuously deforming them, see videos in the companion repository \href{https://github.com/KZL358/Polarisation-Singularities}{\textcolor{black}{\faGithub}} \cite{figshare}.

\paragraph*{Line length density:}
We computed the length density of C lines, corresponding to the sum of the lengths of C-lines per unit volume, for both electromagnetic and gravitational waves. After approximately $50$ plane waves, the density converges; we therefore fix the number of plane waves and examine convergence with respect to progressively finer grids. We find that the electromagnetic C-line density converges to $(8.334 \pm 0.018)\,\lambda^{-2}$ and the corresponding gravitational-wave density to $(8.235 \pm 0.008)\,\lambda^{-2}$, where the quoted uncertainties represent the standard errors ($95\%$ confidence intervals). Remarkably, the electromagnetic result agrees exactly with the theoretical value derived in \cite{Berry2001}
\begin{equation}
    d_\text{C}=k^2\Big(\frac{3}{10\pi}+\frac{1}{5\sqrt{3}}\Big)\approx8.3283\lambda^{-2}\;.
\end{equation}%
No analogous computation for gravitational waves currently exists; we plan to address this in future work.
    
\section{Discussion and conclusions}
Generic gravitational wave fields are expected to be threaded by a complex network of polarisation singularities. In monochromatic components of gravitational wave superpositions, circular-polarisation singularities (C lines) form continuous filaments in space that can close into loops and, in general, realise nontrivial linked configurations, while for spin-2 waves linear-polarisation singularities occur as isolated L points.

The main goal of this work was to extend the concept of polarisation singularities, well developed for electromagnetic waves, to higher-spin fields, and in particular to gravitational waves. We have shown that the formalism can be generalised in a consistent and geometrically meaningful way, and that polarisation singularities are a generic feature of wave fields independently of their spin. These objects are topologically stable: C lines and L points are defined locally, with C lines as loci where the energy density is constant in time and L points as zeros of the spin density, even for a spin-$n$ wave. As a result, they cannot be removed by small perturbations but can only move continuously or, in the case of L points, be created or annihilated in opposite-sign pairs \cite{Vernon2022,Vernon:23}.

This topological robustness makes them relevant for gravitational-wave phenomenology whenever the field cannot be described by a single plane wave. This includes the stochastic gravitational-wave background \cite{PhysRevD.99.023534,LISACosmologyWorkingGroup:2022kbp,Mentasti:2023gmg} (see e.g.\ \cite{Caprini:2018mtu} for a review), possible interference from simultaneous arrivals of (quasi-)monochromatic $N>3$ black hole mergers, and continuous, quasi-monochromatic signals from pulsars \cite{galaxies10030072,Moragues:2022aaf}. Indirect effects may also arise through parity-sensitive observables, for example via imprints on the polarisation of the Cosmic Microwave Background\cite{Garcia-Cely:2025mgu}. In all these cases, the singularity network would most naturally enter through statistical properties of the field.

Our analysis clarifies how the structure of these singularities depends on spin. The codimension of circular polarisation singularities is spin-independent, so C-singularities form lines for all spins. By contrast, the codimension of linear polarisation singularities depends on the spin: they are L lines for spin-1 waves but L points for spin-2 waves, and we showed this behaviour to extend to higher spins.

Overall, the present work shows that the topology of the wave field itself, not only its sources, provides a natural and robust structure that may play a role in the phenomenology of gravitational radiation. This may important consequences for the Cosmic Microwave Background and Stochastic Gravitational Wave Background. We hope to report on some of these points in the future.

\acknowledgments
We are grateful to Henri Inchauspé for useful comments and discussions.
C.R. acknowledges support from the
Science and Technology Facilities Council (STFC). SG and FJRF acknowledge support from the EIC Pathfinder project CHIRALFORCE (Grant No.~101046961), funded by the Innovate UK Horizon Europe Guarantee (UKRI Project No.~10045438). K.L acknowledges support from the Engineering and Physical
Sciences Research Council (EPSRC), grant number
EP/Y015673. 

\appendix
\crefalias{section}{appendix}
\section{Codimension of polarisation singularities}\label{app_codimension}
Let us justify the codimension of the polarisation singularity. As reminded by \cite{Bliokh2021}, \textit{``The codimension of a
    geometric object is the difference between the dimension of
    the space in which it lies and its own dimension"}. In the case of C lines, for example, we have two conditions expressed on the vector $\mathbf{E}$, but they can be translated to conditions on the coordinate space. This is because the vector $\mathbf{E}$ itself is a function of the coordinates. For example, in Cartesian coordinates, we can decompose it: 
\begin{equation*}
    \,\mathbf{E}(x,y,z)= E_x(x,y,z)\hat{\mathbf{e}}_x+E_y(x,y,z) \hat{\mathbf{e}}_y+E_z(x,y,z)\hat{\mathbf{e}}_z.\!
\end{equation*}
So if we look at the two conditions given by setting \cref{psi_EM} to zero, they translate to: 
\begin{equation}
\begin{cases}
\operatorname{Re}\!\left[\mathbf{E}(x,y,z)\cdot \mathbf{E}(x,y,z)\right]=F_1(x,y,z)=0\;,\\
\operatorname{Im}\!\left[\mathbf{E}(x,y,z)\cdot \mathbf{E}(x,y,z)\right]=F_2(x,y,z)=0\;.
\end{cases}
\end{equation}
Therefore, we have $2$ conditions in real space for C lines, and we are considering a $3$D space. Thus, the codimension of C lines is $3-2=1$. This justifies why they are lines rather than surfaces or points.

\section{Polarisation singularities for spin-s fields}\label{app:higher_spin}
We analyse the codimension of C and L polarisation singularities for monochromatic spin-$s$ waves, with $s$ a non-negative integer. A real spin-$s$ gauge field can be represented as a completely symmetric and traceless tensor, corresponding to an irreducible representation of $\mathrm{SO}(3)$,
\begin{equation}
    \mathcal{A}_{i_1\ldots i_s}(\mathbf{r}, t)=\operatorname{Re}[A_{i_1\ldots i_s}(\mathbf{r})e^{-i\omega t}]\;.
\end{equation}
For such a field, the local energy density, $u_A$, and spin angular momentum density, $\mathbf{S}_A$, are proportional to \cite{Maggiore2008}:
\begin{equation}
    \!\!u_A\propto\dot{\mathcal{A}}_{i_1\ldots i_s}\dot{\mathcal{A}}^{i_1\ldots i_s},\quad\!\!
    \mathbf{S}_A\propto\varepsilon^{ijk}{\mathcal{A}}_{ji_2\ldots i_s}\dot{\mathcal{A}}_k{}^{i_2\ldots i_s}\hat{\mathbf{e}}_i\,,\!\!
\end{equation}
where the overdot denotes a time derivative. For time-harmonic fields, the instantaneous energy density can be written as a sum of a constant and oscillatory contributions as follows:
\begin{equation}
    \!\!u_A\propto\tfrac{\omega^2}{2}\!\big[{{A}}^*_{i_1\ldots i_s}{{A}}^{i_1\ldots i_s}+\operatorname{Re}(\underbrace{{{A}}_{i_1\ldots i_s}{{A}}^{i_1\ldots i_s}}_{\psi_A}e^{-2i\omega t})\big].\!\!
\end{equation}
The complex scalar $\psi_A$ generalises the familiar quantity that vanishes at C points for vector waves. Its vanishing implies that the energy density is strictly constant at those points.

Similarly, the spin angular momentum density is
\begin{equation}
    \mathbf{S}_A\propto\omega\underbrace{\tfrac{1}{2}\epsilon^{ijk} \operatorname{Im}(A^*_{j i_2 \ldots i_s} A_{k}{}^{i_2 \ldots i_s}) \,\hat{\mathbf{e}}_i}_{\mathbf{n}_A}\;,
\end{equation}
where $\mathbf n_A$ is the natural generalisation of the polarisation ellipse normal vector. Its vanishing defines L singularities, which are characterised by the vanishing of the spin angular momentum density.

\paragraph*{C lines} Independently of the spin $s$, the condition $\psi_A=0$ imposes two real constraints, corresponding to the vanishing of its real and imaginary parts. In three-dimensional space, this generically leads to one-dimensional singular sets, and hence C lines persist for arbitrary spin.

\paragraph*{L points} The situation is different for the condition $\mathbf n_A=\mathbf 0$. Writing
\(
A_{i_1\ldots i_s}=P_{i_1\ldots i_s}+iQ_{i_1\ldots i_s}
\),
this condition corresponds to three real equations,
\begin{equation*}
\begin{cases}
P_{x i_2\ldots i_s} Q_{y}{}^{i_2\ldots i_s}
=
P_{y i_2\ldots i_s} Q_{x}{}^{i_2\ldots i_s},\\[4pt]
P_{x i_2\ldots i_s} Q_{z}{}^{i_2\ldots i_s}
=
P_{z i_2\ldots i_s} Q_{x}{}^{i_2\ldots i_s},\\[4pt]
P_{z i_2\ldots i_s} Q_{y}{}^{i_2\ldots i_s}
=
P_{y i_2\ldots i_s} Q_{z}{}^{i_2\ldots i_s}.
\end{cases}
\end{equation*}
For a generic integer spin-$s>1$ field, these are independent constraints, implying isolated L points in three-dimensional space.

In contrast, for $s=1$ the index contractions disappear, and the three conditions reduce to
\begin{equation*}
    \frac{P_x}{Q_x}=\frac{P_y}{Q_y}=\frac{P_z}{Q_z},
\end{equation*}
which represents only two independent constraints. As a result, vector waves generically exhibit L lines rather than isolated points.

Finally, the definitions of C and L singularities in terms of the energy density $u_A$ and spin density $\mathbf S_A$ remain consistent even for a real scalar field ($s=0$). In this case, the spin density vanishes identically, so all of space constitutes an L volume. The C condition $\psi_A=A^2=0$ is satisfied only at the nodes $A(\mathbf r)=0$, yielding nodal lines. This is consistent with the general statement that points satisfying both $\psi_A=0$ and $\mathbf n_A=\mathbf 0$ must correspond to the vanishing of the field,
\(
A_{i_1\ldots i_s}(\mathbf r)=0.
\) 

\section{Visualising polarisation singularities}\label{visualisation}
Unlike in electromagnetism, the fields $P_{ij}$ and $Q_{ij}$ are tensorial rather than vectorial, which makes the local polarisation state at a point $\mathbf{r}$ less straightforward to visualise. A standard approach is to examine the tidal motion of freely falling test masses via the geodesic deviation equation. To do so, imagine test masses distributed isotropically around the observation point, placed on a unit sphere with directions $\hat{\mathbf{n}}$. 
As a gravitational wave passes, each mass acquires a small displacement $\delta \mathbf{n}$, satisfying the geodesic deviation equation, $\delta \ddot{\mathbf{n}}=\tfrac12\ddot{\mathcal{h}}^i{}_{j}\hat{n}^j\hat{\mathbf{e}}_i$, where the overdot denotes a time derivative, which produces a vector field over the sphere that encodes the local tidal deformation:
\begin{equation}
\begin{split}
    \delta \mathbf{n}(\mathbf{r}, t)&=\tfrac12\mathcal{h}^i{}_{j}(\mathbf{r}, t)\hat{n}^j\hat{\mathbf{e}}_i\\
    &=\tfrac12[P^i{}_{j}(\mathbf{r})\cos{\omega t}+Q^i{}_{j}(\mathbf{r})\sin{\omega t}]\hat{n}^j\hat{\mathbf{e}}_i\;.
\end{split}
\end{equation}
For each line of sight $\hat{n}^{i}$, the trajectory of $\delta \mathbf{n}(t)$ lies
in the two-dimensional subspace spanned by $P^{i}{}_{j}\hat{n}^{j}\hat{\mathbf{e}}_i$ and
$Q^{i}{}_{j}\hat{n}^{j}\hat{\mathbf{e}}_i$, and in general, it traces an ellipse in that plane.
Since $\delta \mathbf{n}(t)$ is a time-harmonic vector field, we can define the normal to this ellipse exactly as in \eqref{normal}:
\begin{equation}\label{normal_dn}
    \mathbf{n}_{\delta \mathbf{n}}=\tfrac14(\varepsilon^{ijk}P_{jl}Q_{km})\hat{n}^l\hat{n}^m\hat{\mathbf{e}}_i\;,
\end{equation}
together with the complex scalar analogous to \eqref{psi_EM},
\begin{equation}\label{psi_dn}
    \psi_{\delta \mathbf{n}}=\tfrac14(h^i{}_kh_{il})\hat{n}^k\hat{n}^l\;.
\end{equation}
The ellipses traced by $\delta \mathbf{n}(t)$ vary with direction $\hat{\mathbf{n}}$, but since \cref{normal_dn,psi_dn} are quadratic in $\hat{\mathbf{n}}$, we can perform the orientation average over the unit sphere: \[\langle{A}\rangle=\tfrac{1}{4\pi}\iint_{S^2} A(\hat{\mathbf{n}}) \mathrm{d}{\Omega}\,.\] Using the identity $\langle\hat{n}^i\hat{n}^j\rangle=\delta^{ij}/3$, we obtain, up to a factor, exactly the normal and the complex scalar in \cref{normal_GR,psi_GR}
\begin{equation}\label{eval}
    \langle \mathbf{n}_{\delta \mathbf{n}}\rangle=\tfrac1{12}\mathbf{n}_{h}\,,\quad 
    \langle\psi_{\delta \mathbf{n}}\rangle=\tfrac1{12}\psi_{h}\,.
\end{equation}
This implies that gravitational singularities are averages and that the test masses, in general, still move along ellipses. In order for all test masses to move along circular or linear trajectories, we would require a stronger condition, \cref{psi_dn,normal_dn} vanishing for any $\hat{\mathbf{n}}$:
\[h^i{}_kh_{il}=0\,, \quad \text{or}\quad\varepsilon^{ijk}h_{jl}^*h_{km}=0\,,\]
which will not happen generically, as that would impose more than three constraints.

\section{Plane wave polarisation basis for a general propagation direction}\label{app_extension}
In order to produce the \cref{interference}, we need to add $N$ plane waves with random propagation directions $\smash{\hat{\mathbf{k}}_n}=\mathbf{k}_n/k$, where $k=\omega/c$, and with random complex amplitudes ($E_n^\theta$, $E_n^\varphi$ or $h_n^+$, $h_n^\times$). The {resulting} electric field is: 
\begin{equation}
    E_i({\mathbf{r}})=\sum_{n=1}^N[E_n^\theta \hat{e}^\theta_i(\mathbf{k}_n)+E_n^\varphi \hat{e}^\varphi_i(\mathbf{k}_n)]e^{i\mathbf{k}_n\cdot\mathbf{r}},
\end{equation}
where the ordered set $\{\hat{\mathbf{e}}_r(\mathbf{k}),\hat{\mathbf{e}}_\theta(\mathbf{k}),\hat{\mathbf{e}}_\varphi(\mathbf{k})\}$, with radial direction defined by the wavevector as $\hat{\mathbf{e}}_r(\mathbf{k})=\hat{\mathbf{k}}$, forms an orthonormal right-handed vector frame, 
\begin{equation}
    \hat{\mathbf{e}}_i(\mathbf{k})\cdot\hat{\mathbf{e}}_j(\mathbf{k})=\delta_{ij}\,,\quad 
    [\hat{\mathbf{e}}_i(\mathbf{k})\times\hat{\mathbf{e}}_j(\mathbf{k})]\cdot\hat{\mathbf{e}}_k(\mathbf{k})=\varepsilon_{ijk}\;,
\end{equation}
 with $\varepsilon_{r\theta\varphi}=1$. This frame can be constructed as:
\begin{equation}
        \hat{\mathbf{e}}_\varphi(\mathbf{k})=\frac{\hat{\mathbf{e}}_z\times\hat{\mathbf{k}}}{||\hat{\mathbf{e}}_z\times\hat{\mathbf{k}}||}\;,\quad
        \hat{\mathbf{e}}_\theta(\mathbf{k})=\hat{\mathbf{k}}\times\hat{\mathbf{e}}_\varphi(\mathbf{k})\;.
\end{equation}
There is an ambiguity when $\hat{\mathbf{k}}=\pm\hat{\mathbf{e}}_z$ due to the hairy ball theorem \cite[p.~423]{FrankelGeometryPhysics}, which forbids the existence of a globally smooth, nonvanishing tangent vector field on the sphere. In that case, we can take the set $\{\pm\hat{\mathbf{e}}_x,\hat{\mathbf{e}}_y,\pm\hat{\mathbf{e}}_z\}$, which avoids the singularity at the poles. Similarly, the gravitational tensor potential in the TT gauge for $N$ plane waves is:
\begin{equation}
    \!\!h_{ij}({\mathbf{r}})=\sum_{n=1}^N[h_n^+\hat{e}^+_{ij}(\mathbf{k}_n)+h_n^\times \hat{e}^\times_{ij}(\mathbf{k}_n)]e^{i\mathbf{k}_n\cdot\mathbf{r}},\!\!
\end{equation}
where we define the two basis tensors as follows:
\begin{align}
\hat e^{+}_{ij}(\mathbf{k})
&=\tfrac{1}{\sqrt{2}}[
\hat e^{\theta}_i(\mathbf{k})\,\hat e^{\theta}_j(\mathbf{k})
-
\hat e^{\varphi}_i(\mathbf{k})\,\hat e^{\varphi}_j(\mathbf{k})]\;,
\\[4pt]
\hat e^{\times}_{ij}(\mathbf{k})
&=\tfrac{1}{\sqrt{2}}[
\hat e^{\theta}_i(\mathbf{k})\,\hat e^{\varphi}_j(\mathbf{k})
+
\hat e^{\varphi}_i(\mathbf{k})\,\hat e^{\theta}_j(\mathbf{k})]\;,
\end{align}
such that they are orthonormal $\hat e^{P}_{ij}(\mathbf{k})\hat e_{P'}^{ij}(\mathbf{k})=\delta^P_{P'}$. Note that this definition differs from the more conventional choice, $\hat e^{P}_{ij}(\mathbf{k})\hat e_{P'}^{ij}(\mathbf{k})=2\delta^P_{P'}$, by a factor of $\sqrt{2}$. The ambiguity at poles can be resolved again, as before, by requiring that $\{\hat{\mathbf{e}}_\theta(\pm\hat{\mathbf{e}}_z),\hat{\mathbf{e}}_\varphi(\pm\hat{\mathbf{e}}_z)\}=\{\pm\hat{\mathbf{e}}_x,\hat{\mathbf{e}}_y\}$.
One can also use the circular polarisation bases:
\begin{equation}
    \!\!A_{I}({\mathbf{r}})=\sum_{n=1}^N[a_n^R\hat{e}^R_{I}(\mathbf{k}_n)+a_n^L \hat{e}^L_{I}(\mathbf{k}_n)]e^{i\mathbf{k}_n\cdot\mathbf{r}},\!\!
\end{equation}
where $A_I({\mathbf{r}})$ can be either the vector $E_i({\mathbf{r}})$ or tensor $h_{ij}({\mathbf{r}})$, with $I$ representing one ($i$) or two ($ij$) indices, $a_n^{R/L}$ are complex amplitudes, and $\hat{e}^{R/L}_{I}(\mathbf{k})$ are
\begin{equation}
    \begin{split}
        \hat{e}^{R/L}_{i}(\mathbf{k})&=\tfrac{1}{\sqrt{2}}[\hat{e}^\theta_i(\mathbf{k})\pm i\hat{e}^\phi_i(\mathbf{k})]\;,\\
        \hat{e}^{R/L}_{ij}(\mathbf{k})&=\tfrac{1}{\sqrt{2}}[\hat{e}^+_{ij}(\mathbf{k})\pm i\hat{e}^\times_{ij}(\mathbf{k})]\;,
    \end{split}
\end{equation}
which are the spin-1 and spin-2 orthonormal circular bases, respectively. 

\bibliography{main}

@article{LIGO,
	author = "Abbott, B. P. and others",
	collaboration = "{LIGO Scientific and Virgo Collaborations}",
	title = "{Observation of Gravitational Waves from a Binary Black Hole Merger}",
	reportNumber = "LIGO-P150914",
	doi = "10.1103/PhysRevLett.116.061102",
	journal = "Phys. Rev. Lett.",
	volume = "116",
	number = "6",
	pages = "061102",
	year = "2016"
}

@article{LIGOScientific:2017ycc,
    author = "Abbott, B. P. and others",
  collaboration  = {LIGO Scientific and Virgo Collaborations},
    title = "{GW170814: A Three-Detector Observation of Gravitational Waves from a Binary Black Hole Coalescence}",
    doi = "10.1103/PhysRevLett.119.141101",
    journal = "Phys. Rev. Lett.",
    volume = "119",
    number = "14",
    pages = "141101",
    year = "2017"
}

@article{LIGOScientific:2020kqk,
    author = "Abbott, R. and others",
    collaboration  = {LIGO Scientific and Virgo Collaborations},
    title = "{Population Properties of Compact Objects from the Second LIGO-Virgo Gravitational-Wave Transient Catalog}",
    reportNumber = "LIGO-P2000077",
    doi = "10.3847/2041-8213/abe949",
    journal = "Astrophys. J. Lett.",
    volume = "913",
    number = "1",
    pages = "L7",
    year = "2021"
}

@article{KAGRA:2021vkt,
    author = "Abbott, R. and others",
    collaboration  = {LIGO Scientific, Virgo, and KAGRA Collaborations},
    title = "{GWTC-3: Compact Binary Coalescences Observed by LIGO and Virgo during the Second Part of the Third Observing Run}",
    reportNumber = "LIGO-P2000318",
    doi = "10.1103/PhysRevX.13.041039",
    journal = "Phys. Rev. X",
    volume = "13",
    number = "4",
    pages = "041039",
    year = "2023"
}

@article{KAGRA:2021duu,
    author = "Abbott, R. and others",
    collaboration  = {LIGO Scientific, Virgo, and KAGRA Collaborations},
    title = "{Population of Merging Compact Binaries Inferred Using Gravitational Waves through GWTC-3}",
    reportNumber = "LIGO-P2100239 ; Data release: https://zenodo.org/record/5655785, LIGO-P2100239",
    doi = "10.1103/PhysRevX.13.011048",
    journal = "Phys. Rev. X",
    volume = "13",
    number = "1",
    pages = "011048",
    year = "2023"
}

@book{Andrews,
  title        = {The Angular Momentum of Light},
  author       = {David L. Andrews and Mohamed Babiker},
  year         = {2012},
  publisher    = {Cambridge University Press},
  doi          = {10.1017/CBO9780511795213},
}

@Article{Barnett2014,
  author    = {Stephen M Barnett},
  title     = {Maxwellian theory of gravitational waves and their mechanical properties},
  doi       = {10.1088/1367-2630/16/2/023027},
  number    = {2},
  pages     = {023027},
  volume    = {16},
  fjournal  = {New Journal of Physics},
  journal   = {New J. Phys.},
  month     = feb,
  publisher = {IOP Publishing},
  year      = {2014},
}

@article{LISACosmologyWorkingGroup:2022kbp,
    author        = {Bartolo, Nicola and others},
    collaboration = {LISA Cosmology Working Group},
    title         = {Probing anisotropies of the Stochastic Gravitational Wave Background with LISA},
    journal       = {JCAP},
    volume        = {2022},
    number        = {11},
    pages         = {009},
    year          = {2022},
    doi           = {10.1088/1475-7516/2022/11/009}
}

@Article{Bauer2015,
  author  = {Thomas Bauer and Peter Banzer and Ebrahim Karimi and Sergej Orlov and Andrea Rubano and Lorenzo Marrucci and Enrico Santamato and Robert W. Boyd and Gerd Leuchs},
  title   = {Observation of optical polarization M\"obius strips},
  doi     = {10.1126/science.1260635},
  number  = {6225},
  pages   = {964--966},
  volume  = {347},
  journal = {Science},
  year    = {2015},
}

@Article{PhysRevLett.117.013601,
  author    = {Bauer, Thomas and Neugebauer, Martin and Leuchs, Gerd and Banzer, Peter},
  title     = {Optical Polarization M\"obius Strips and Points of Purely Transverse Spin Density},
  doi       = {10.1103/PhysRevLett.117.013601},
  issue     = {1},
  pages     = {013601},
  volume    = {117},
  journal   = {Phys. Rev. Lett.},
  month     = jun,
  numpages  = {6},
  publisher = {American Physical Society},
  year      = {2016},
}

@Article{Berry2001,
  author   = {Berry, M.V and Dennis, M.R},
  title    = {Polarization singularities in isotropic random vector waves},
  doi      = {10.1098/rspa.2000.0660},
  number   = {2005},
  pages    = {141--155},
  volume   = {457},
  fjournal = {Proceedings of the Royal Society of London. Series A: Mathematical, Physical and Engineering Sciences},
  journal  = {Proc. R. Soc. Lond. A},
  year     = {2001},
}

@Article{Bliokh2021,
  author   = {Bliokh, Konstantin Y. and Alonso, Miguel A. and Sugic, Danica and Perrin, Mathias and Nori, Franco and Brasselet, Etienne},
  title    = {{Polarization singularities and Möbius strips in sound and water-surface waves}},
  doi      = {10.1063/5.0056333},
  issn     = {1070-6631},
  number   = {7},
  pages    = {077122},
  volume   = {33},
  fjournal = {Physics of Fluids},
  journal  = {Phys. Fluids},
  month    = jul,
  year     = {2021},
}

@Article{PhysRevLett.102.033902,
  author    = {Burresi, M. and Engelen, R. J. P. and Opheij, A. and van Oosten, D. and Mori, D. and Baba, T. and Kuipers, L.},
  title     = {Observation of Polarization Singularities at the Nanoscale},
  doi       = {10.1103/PhysRevLett.102.033902},
  issue     = {3},
  pages     = {033902},
  volume    = {102},
  journal   = {Phys. Rev. Lett.},
  month     = jan,
  numpages  = {4},
  publisher = {American Physical Society},
  year      = {2009},
}

@misc{GRACE,
    key  = {LIGO Scientific, Virgo, and KAGRA Collaborations},
    collaboration  = {LIGO Scientific, Virgo, and KAGRA Collaborations},
    title        = {{GraceDB} {LVK} Public Alerts},
    url          = {https://gracedb.ligo.org/superevents/public/O4/},
    note         = {Online database},
    year         = {2025}
}

@Article{PhysRevD.99.023534,
  author    = {Cusin, Giulia and Durrer, Ruth and Ferreira, Pedro G.},
  title     = {Polarization of a stochastic gravitational wave background through diffusion by massive structures},
  doi       = {10.1103/PhysRevD.99.023534},
  issue     = {2},
  pages     = {023534},
  volume    = {99},
  journal   = {Phys. Rev. D},
  month     = jan,
  numpages  = {26},
  publisher = {American Physical Society},
  year      = {2019},
}

@Article{Knots,
  author   = {{Dennis}, Mark R. and {King}, Robert P. and {Jack}, Barry and {O'Holleran}, Kevin and {Padgett}, Miles J.},
  title    = {{Isolated optical vortex knots}},
  doi      = {10.1038/nphys1504},
  number   = {2},
  pages    = {118--121},
  volume   = {6},
  adsurl   = {https://ui.adsabs.harvard.edu/abs/2010NatPh...6..118D},
  fjournal = {Nature Physics},
  journal  = {Nat. Phys.},
  month    = feb,
  year     = {2010},
}

@Article{FREUND20101,
  author   = {Isaac Freund},
  title    = {Optical Möbius strips in three-dimensional ellipse fields: I. Lines of circular polarization},
  doi      = {10.1016/j.optcom.2009.09.042},
  issn     = {0030-4018},
  number   = {1},
  pages    = {1--15},
  url      = {https://www.sciencedirect.com/science/article/pii/S0030401809009195},
  volume   = {283},
  fjournal = {Optics Communications},
  journal  = {Opt. Commun.},
  year     = {2010},
}

@article{Galaudage:2021rkt,
    author = "Galaudage, Shanika and others",
    title = "{Building Better Spin Models for Merging Binary Black Holes: Evidence for Nonspinning and Rapidly Spinning Nearly Aligned Subpopulations}",
    doi = "10.3847/2041-8213/ac2f3c",
    journal = "Astrophys. J. Lett.",
    volume = "921",
    number = "1",
    pages = "L15",
    year = "2021",
}

@Article{Galvez2017,
  author   = {Galvez, Enrique and Dutta, Ishir and Beach, Kory and Zeosky, Jon and Jones, Joshua and Khajavi, Behzad},
  title    = {Multitwist M{\"o}bius Strips and Twisted Ribbons in the Polarization of Paraxial Light Beams},
  doi      = {10.1038/s41598-017-13199-1},
  pages    = {13199},
  volume   = {7},
  fjournal = {Scientific Reports},
  journal  = {Sci. Rep.},
  month    = oct,
  year     = {2017},
}

@misc{figshare,
  author       = {Kyan Louisia and Alex J. Vernon  and Claire Rigouzzo and Sebastian Golat},
  title        = {{MATLAB} code for numerical simulations and data analysis used in this work},
  URL          = {https://github.com/KZL358/Polarisation-Singularities},
  note         = {{GitHub} repository},
  year         = {2025},
}

@Article{PhysRevLett.111.150404,
  author    = {Kedia, Hridesh and Bialynicki-Birula, Iwo and Peralta-Salas, Daniel and Irvine, William T. M.},
  title     = {Tying Knots in Light Fields},
  doi       = {10.1103/PhysRevLett.111.150404},
  issue     = {15},
  pages     = {150404},
  volume    = {111},
  journal   = {Phys. Rev. Lett.},
  month     = oct,
  numpages  = {5},
  publisher = {American Physical Society},
  year      = {2013},
}

@article{2015PhRvA.92f3819L,
	author = {{Lang}, Ben and {Beggs}, Daryl M. and {Young}, Andrew B. and {Rarity}, John G. and {Oulton}, Ruth},
	title = {Stability of polarization singularities in disordered photonic crystal waveguides},
	journal = {\pra},
	year = 2015,
	month = dec,
	volume = {92},
	number = {6},
	eid = {063819},
	pages = {063819},
	doi = {10.1103/PhysRevA.92.063819},
}

@Article{e6e9247d3ab54f1a80459d871decfe3a,
  author    = {J Leach and MR Dennis and J Courtial and MJ Padgett},
  title     = {Knotted threads of darkness},
  doi       = {10.1038/432165a},
  issn      = {1476-4687},
  note      = {Publisher: Nature Publishing Group},
  pages     = {165--165},
  volume    = {432 (7014)},
  journal   = {Nature},
  month     = nov,
  publisher = {Springer Nature},
  year      = {2004},
}

@book{FrankelGeometryPhysics,
  author    = {Frankel, Theodore},
  title     = {The Geometry of Physics: An Introduction},
  edition   = {3},
  publisher = {Cambridge University Press},
  address   = {Cambridge},
  year      = {2012},
  doi       = {https://doi.org/10.1017/CBO9781139061377},
  isbn      = {9781139061377}
}

@article{Mandel:2018hfr,
    author = "Mandel, Ilya and Farmer, Alison",
    title = "{Merging stellar-mass binary black holes}",
    doi = "10.1016/j.physrep.2022.01.003",
    journal = "Phys. Rept.",
    volume = "955",
    pages = "1--24",
    year = "2022"
}

@article{Mentasti:2023gmg,
    author  = {Mentasti, Giorgio and Contaldi, Carlo and Peloso, Marco},
    title   = {Prospects for detecting anisotropies and polarization of the stochastic gravitational wave background with ground-based detectors},
    journal = {JCAP},
    volume  = {2023},
    number  = {08},
    pages   = {053},
    year    = {2023},
    doi     = {10.1088/1475-7516/2023/08/053}
}

@Article{Muelas_Hurtado_2022,
  author    = {Muelas-Hurtado, Ruben D. and Volke-Sepúlveda, Karen and Ealo, Joao L. and Nori, Franco and Alonso, Miguel A. and Bliokh, Konstantin Y. and Brasselet, Etienne},
  title     = {Observation of Polarization Singularities and Topological Textures in Sound Waves},
  doi       = {10.1103/physrevlett.129.204301},
  issn      = {1079-7114},
  number    = {20},
  pages     = {204301},
  volume    = {129},
  fjournal  = {Physical Review Letters},
  journal   = {Phys. Rev. Lett.},
  month     = nov,
  publisher = {American Physical Society (APS)},
  year      = {2022},
}

@Article{Nye1987,
  author   = {Nye, John Frederick and Hajnal, J. V.},
  title    = {The wave structure of monochromatic electromagnetic radiation},
  doi      = {10.1098/rspa.1987.0002},
  number   = {1836},
  pages    = {21--36},
  volume   = {409},
  fjournal = {Proceedings of the Royal Society of London. Series A: Mathematical, Physical and Engineering Sciences},
  journal  = {Proc. R. Soc. Lond. A},
  year     = {1987},
}

@Article{Pisanty_2019,
  author    = {Pisanty, Emilio and Machado, Gerard J. and Vicuña-Hernández, Verónica and Picón, Antonio and Celi, Alessio and Torres, Juan P. and Lewenstein, Maciej},
  title     = {Knotting fractional-order knots with the polarization state of light},
  doi       = {10.1038/s41566-019-0450-2},
  issn      = {1749-4893},
  number    = {8},
  pages     = {569--574},
  volume    = {13},
  fjournal  = {Nature Photonics},
  journal   = {Nat. Photonics},
  month     = jun,
  publisher = {Springer Science and Business Media LLC},
  year      = {2019},
}

@Article{Spaegele2023,
  author   = {Christina M. Spaegele and Michele Tamagnone and Soon Wei Daniel Lim and Marcus Ossiander and Maryna L. Meretska and Federico Capasso},
  title    = {Topologically protected optical polarization singularities in four-dimensional space},
  doi      = {10.1126/sciadv.adh0369},
  number   = {24},
  pages    = {eadh0369},
  volume   = {9},
  fjournal = {Science Advances},
  journal  = {Sci. Adv.},
  year     = {2023},
}

@Book{Torres,
  author    = {Torres, Juan and Torner, L.},
  title     = {Twisted Photons: Applications of Light with Orbital Angular Momentum},
  doi       = {10.1002/9783527635368},
  isbn      = {9783527409075},
  publisher = {Wiley-VCH},
  month     = mar,
  year      = {2011},
}

@Article{Wang2012TerabitFD,
  author   = {Jian Wang and Jeng-Yuan Yang and Irfan Fazal and Nisar Ahmed and Yan Yan and Hao Huang and Yongxiong Ren and Yang Yue and Samuel Dolinar and Moshe Tur and Alan E. Willner},
  title    = {Terabit free-space data transmission employing orbital angular momentum multiplexing},
  pages    = {488--496},
  url      = {https://api.semanticscholar.org/CorpusID:5914771},
  volume   = {6},
  fjournal = {Nature Photonics},
  journal  = {Nat. Photonics},
  year     = {2012},
}

@article{Caprini:2018mtu,
    author = "Caprini, Chiara and Figueroa, Daniel G.",
    title = "{Cosmological Backgrounds of Gravitational Waves}",
    doi = "10.1088/1361-6382/aac608",
    journal = "Class. Quant. Grav.",
    volume = "35",
    number = "16",
    pages = "163001",
    year = "2018"
}

@article{Garcia-Cely:2025mgu,
    author = "Garc{\'\i}a-Cely, Camilo and Marsili, Luca and Ringwald, Andreas and Spector, Aaron D.",
    title = "{Polarimetric searches for axion dark matter and high-frequency gravitational waves using optical cavities}",
    doi = "10.1103/26vg-wcfx",
    journal = "Phys. Rev. D",
    volume = "112",
    number = "2",
    pages = "023031",
    year = "2025"
}

@article{Moragues:2022aaf,
    author = "Moragues, Joan and Modafferi, Luana M. and Tenorio, Rodrigo and Keitel, David",
    title = "{Prospects for detecting transient quasi-monochromatic gravitational waves from glitching pulsars with current and future detectors}",
    reportNumber = "LIGO-P2200273 ET-0198D-22, LIGO-P2200273 ET-0198A-22",
    doi = "10.1093/mnras/stac3665",
    journal = "Mon. Not. Roy. Astron. Soc.",
    volume = "519",
    number = "4",
    pages = "5161--5176",
    year = "2023"
}

@Article{galaxies10030072,
AUTHOR = {Piccinni, Ornella Juliana},
TITLE = {Status and Perspectives of Continuous Gravitational Wave Searches},
JOURNAL = {Galaxies},
VOLUME = {10},
YEAR = {2022},
NUMBER = {3},
ARTICLE-NUMBER = {72},
URL = {https://www.mdpi.com/2075-4434/10/3/72},
ISSN = {2075-4434},
DOI = {10.3390/galaxies10030072}
}

@article{Golat2025Oct,
	author = {Golat, Sebastian and Vernon, Alex J. and Rodr{\ifmmode\acute{\imath}\else\'{\i}\fi}guez-Fortu{\ifmmode\tilde{n}\else\~{n}\fi}o, Francisco J.},
	title = {{The electromagnetic symmetry sphere: a framework for energy, momentum, spin and other electromagnetic quantities}},
	journal = {Phys. Scr.},
	volume = {100},
	number = {10},
	pages = {105518},
	year = {2025},
	month = oct,
	issn = {1402-4896},
	publisher = {IOP Publishing},
	doi = {10.1088/1402-4896/ae0662}
}

@book{Maggiore2008,
	author = {Maggiore, Michele},
	title = {{Gravitational Waves: Volume 1: Theory and Experiments}},
	year = {2008},
	isbn = {978-0-19857074-5},
	publisher = {Oxford University Press},
	address = {Oxford, England, UK},
	url = {https://books.google.nl/books/about/Gravitational_Waves.html?id=AqVpQgAACAAJ&redir_esc=y}
}

@book{MTW,
  title     = {Gravitation},
  author    = {Misner, Charles W. and Thorne, Kip S. and Wheeler, John Archibald},
  publisher = {W. H. Freeman and Company},
  year      = {1973},
  address   = {San Francisco, CA, USA},
  isbn      = {978-0716703440}
}

@misc{Vernon2025Jul,
	author = {Vernon, Alex J.},
	title = {{Topologies of light in electric-magnetic space}},
    
	year = {2025},
	month = jul,
	eprint = {2507.16721},
    archiveprefix = {arXiv},
}

@article{Vernon:23,
author = {Vernon, Alex J. and Dennis, Mark R. and Rodr\'{i}guez-Fortu\~{n}o, Francisco J. },
journal = {Optica},
number = {9},
pages = {1231--1240},
publisher = {Optica Publishing Group},
title = {{3D} zeros in electromagnetic fields},
volume = {10},
month = {Sep},
year = {2023},
url = {https://opg.optica.org/optica/abstract.cfm?URI=optica-10-9-1231},
doi = {10.1364/OPTICA.487333},
}

@article{Vernon2022,
  author    = {Vernon, Alex J. and Rodr\'{\i}guez-Fortu\~no, Francisco J.},
  title     = {Creating and moving nanoantenna cold spots anywhere},
  journal   = {Light: Sci. Appl.},
  fjournal  = {Light: Science \& Applications},
  year      = {2022},
  volume    = {11},
  number    = {1},
  pages     = {258},
  month     = aug,
  doi       = {10.1038/s41377-022-00893-7},
  url       = {https://doi.org/10.1038/s41377-022-00893-7},
  issn      = {2047-7538}
}

@phdthesis{dennis2001topological,
  title={Topological singularities in wave fields},
  author={Dennis, Mark Richard},
  school={University of Bristol},
  year={2001},
 url={https://www.bristol.ac.uk/physics/media/theory-theses/dennis-mr-thesis.pdf}
}

@article{Rubinsztein-Dunlop2017,
   author = {Halina Rubinsztein-Dunlop and Andrew Forbes and M V Berry and M R Dennis and David L Andrews and Masud Mansuripur and Cornelia Denz and Christina Alpmann and Peter Banzer and Thomas Bauer and Ebrahim Karimi and Lorenzo Marrucci and Miles Padgett and Monika Ritsch-Marte and Natalia M Litchinitser and others},
   doi = {10.1088/2040-8978/19/1/013001},
   issn = {2040-8978},
   issue = {1},
   journal = {Journal of Optics},
   month = {1},
   pages = {013001},
   title = {Roadmap on structured light},
   volume = {19},
   year = {2017}
}

@article{Balzarotti2017,
   author = {Francisco Balzarotti and Yvan Eilers and Klaus C. Gwosch and Arvid H. Gynnå and Volker Westphal and Fernando D. Stefani and Johan Elf and Stefan W. Hell},
   doi = {10.1126/science.aak9913},
   issn = {0036-8075},
   issue = {6325},
   journal = {Science},
   month = {2},
   pages = {606-612},
   title = {Nanometer resolution imaging and tracking of fluorescent molecules with minimal photon fluxes},
   volume = {355},
   year = {2017}
}

@article{Hell1994,
   author = {Stefan W. Hell and Jan Wichmann},
   doi = {10.1364/OL.19.000780},
   issn = {0146-9592},
   issue = {11},
   journal = {Optics Letters},
   month = {6},
   pages = {780},
   title = {Breaking the diffraction resolution limit by stimulated emission: stimulated-emission-depletion fluorescence microscopy},
   volume = {19},
   year = {1994}
}

@article{Nelson2007,
   author = {Karl D. Nelson and Xiao Li and David S. Weiss},
   doi = {10.1038/nphys645},
   issn = {1745-2473},
   issue = {8},
   journal = {Nature Physics},
   month = {8},
   pages = {556-560},
   title = {Imaging single atoms in a three-dimensional array},
   volume = {3},
   year = {2007}
}

@article{Bozinovic2013,
   author = {Nenad Bozinovic and Yang Yue and Yongxiong Ren and Moshe Tur and Poul Kristensen and Hao Huang and Alan E. Willner and Siddharth Ramachandran},
   doi = {10.1126/science.1237861},
   issn = {0036-8075},
   issue = {6140},
   journal = {Science},
   month = {6},
   pages = {1545-1548},
   title = {Terabit-Scale Orbital Angular Momentum Mode Division Multiplexing in Fibers},
   volume = {340},
   year = {2013}
}

@article{Bustamante2021,
   author = {Carlos J. Bustamante and Yann R. Chemla and Shixin Liu and Michelle D. Wang},
   doi = {10.1038/s43586-021-00021-6},
   issn = {2662-8449},
   issue = {1},
   journal = {Nature Reviews Methods Primers},
   month = {3},
   pages = {25},
   title = {Optical tweezers in single-molecule biophysics},
   volume = {1},
   year = {2021}
}

@article{Mayer2024,
   author = {Nicola Mayer and David Ayuso and Piero Decleva and Margarita Khokhlova and Emilio Pisanty and Misha Ivanov and Olga Smirnova},
   doi = {10.1038/s41566-024-01499-8},
   issn = {1749-4885},
   issue = {11},
   journal = {Nature Photonics},
   month = {11},
   pages = {1155-1160},
   title = {Chiral topological light for detection of robust enantiosensitive observables},
   volume = {18},
   year = {2024}
}

@article{Forbes2021,
   author = {Kayn A Forbes and David L Andrews},
   doi = {10.1088/2515-7647/abdb06},
   issn = {2515-7647},
   issue = {2},
   journal = {Journal of Physics: Photonics},
   month = {4},
   pages = {022007},
   title = {Orbital angular momentum of twisted light: chirality and optical activity},
   volume = {3},
   year = {2021}
}

@article{Dorrah2022,
   author = {Ahmed H. Dorrah and Federico Capasso},
   doi = {10.1126/science.abi6860},
   issn = {0036-8075},
   issue = {6591},
   journal = {Science},
   month = {4},
   title = {Tunable structured light with flat optics},
   volume = {376},
   year = {2022}
}

@article{Nye1974,
   author = {John Frederick Nye and Michael Victor Berry},
   doi = {10.1098/rspa.1974.0012},
   issn = {0080-4630},
   issue = {1605},
   journal = {Proceedings of the Royal Society of London. A. Mathematical and Physical Sciences},
   month = {1},
   pages = {165-190},
   title = {Dislocations in wave trains},
   volume = {336},
   year = {1974}
}

@article{Bliokh2023,
   author = {Konstantin Y Bliokh and Ebrahim Karimi and Miles J Padgett and Miguel A Alonso and Mark R Dennis and Angela Dudley and Andrew Forbes and Sina Zahedpour and Scott W Hancock and Howard M Milchberg and Stefan Rotter and Franco Nori and Şahin K Özdemir and Nicholas Bender and Hui Cao and others},
   doi = {10.1088/2040-8986/acea92},
   issn = {2040-8978},
   issue = {10},
   journal = {Journal of Optics},
   month = {10},
   pages = {103001},
   title = {Roadmap on structured waves},
   volume = {25},
   year = {2023}
}
\end{document}